\documentclass{iau307}
\usepackage{graphicx}
\usepackage{natbib}
\usepackage{url}
\usepackage{dtklogos}
\bibpunct{(}{)}{;}{a}{}{,}

\title[Stochastic excitation of gravity waves in rapidly rotating massive stars] 
{Stochastic excitation of gravity waves in rapidly rotating massive stars}

\author[S. Mathis \& C. Neiner]   
{S. Mathis$^{1,2}$
 \and C. Neiner$^2$}

\affiliation{$^1$Laboratoire AIM Paris-Saclay, CEA/DSM - CNRS - Universit\'e Paris Diderot, IRFU/SAp Centre de Saclay, F-91191 Gif-sur-Yvette Cedex, France\\ email: {\tt stephane.mathis@cea.fr} \\[\affilskip]
$^2$LESIA, Observatoire de Paris, CNRS UMR 8109, UPMC, Univ. Paris-Diderot,\\ 5 place Jules Janssen, 92195 Meudon, France}

\pubyear{2014}
\volume{307} 
\pagerange{}
\jname{New windows on massive stars: asteroseismology, interferometry, and spectropolarimetry}
\editors{G. Meynet, C. Georgy, J.H. Groh \& Ph. Stee, eds.}

\begin{document}

\maketitle

\begin{abstract}
Stochastic gravity waves have been recently detected and characterised in stars thanks to space asteroseismology and they may play an important role in the evolution of stellar angular momentum. In this context, the observational study of the CoRoT hot Be star HD\,51452 suggests a potentially strong impact of rotation on stochastic excitation of gravito-inertial waves in rapidly rotating stars. In this work, we present our results on the action of the Coriolis acceleration on stochastic wave excitation by turbulent convection. We study the change of efficiency of this mechanism as a function of the waves' Rossby number and we demonstrate that the excitation presents two different regimes for super-inertial and sub-inertial frequencies. Consequences for rapidly rotating early-type stars and the transport of angular momentum in their interiors are discussed.
\end{abstract}

\section{Stochastic gravity waves in massive stars}

Thanks to asteroseismology using high-resolution photometry in space, our knowledge of stellar structure, rotation, and oscillations is currently undergoing a revolution \citep[e.g.][Aerts, this volume]{ACDK2010}. Until now, our understanding of the excitation of gravity modes in massive stars was mainly based on the $\kappa$-mechanism paradigm. However, these stars have a convective core on the main-sequence as well as a possible sub-surface convective zone that also contribute to waves excitation \citep[e.g.][]{Browningetal2004,Cantielloetal2009, Samadietal2010}. The stochastically excited waves can then reach the surface of the star where they become visible \citep[][]{Shiodeetal2013} as in the case of the CoRoT hot Be star HD\,51452 observed by \cite{Neineretal2012}. Many massive stars are rapidly rotating, as is HD\,51452. Therefore, it becomes important to study the impact of (rapid) rotation on the stochastic excitation of gravity modes, which become gravito-inertial waves (hereafter GIWs) because of the action of the Coriolis acceleration, to provide a correct interpretation of asteroseismic data for such stars. Indeed, both the observation of HD\,51452 and recent numerical simulations \citep{Rogersetal2013} showed a trend in increase of the efficiency of the stochastic excitation mechanism with rotation that must be understood. 

\section{The impact of rotation on stochastic excitation}

\begin{figure}[h!]
\begin{center}
\includegraphics[width=0.5\textwidth]{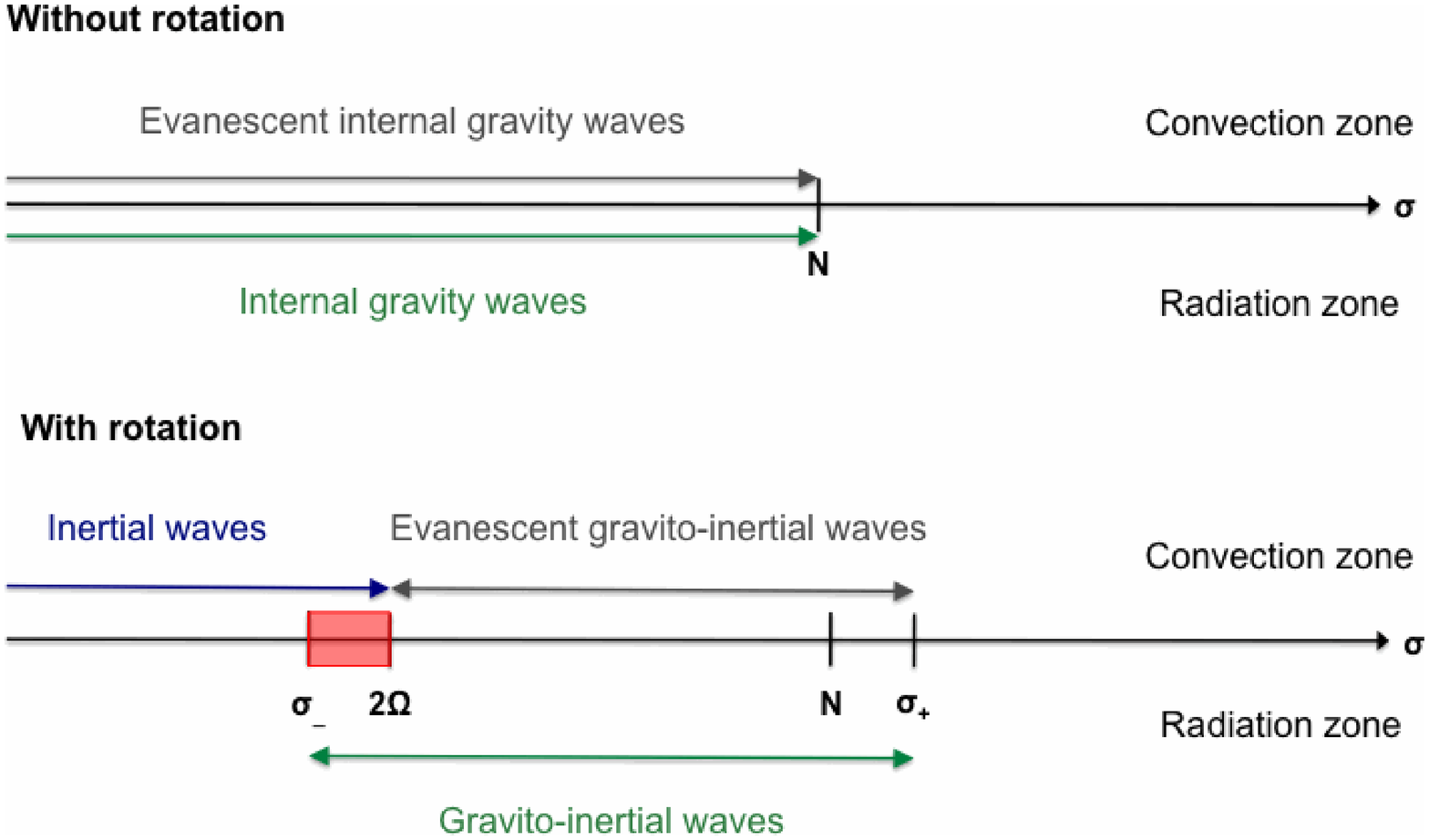} 
\includegraphics[width=0.49\textwidth]{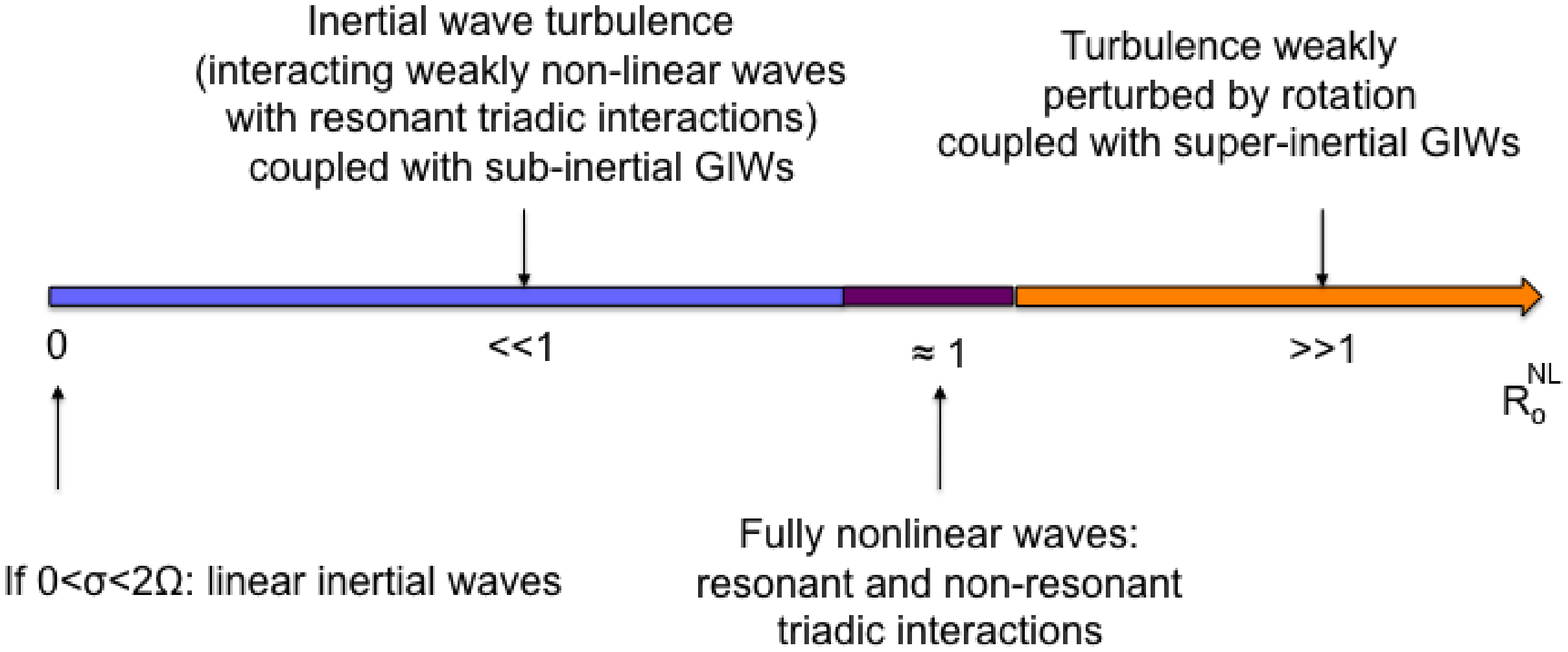} 
\caption{{\bf Left:} low-frequency spectrum of waves in a non-rotating ($\Omega=0$, top) and in a rotating (bottom) star ($N$ is the usual Brunt-V\"ais\"al\"a frequency). {The red box corresponds to sub-inertial GIWs where the cavities of gravity and inertial waves are coupled.} {\bf Right:} wave-turbulence couplings as a function of the non-linear Rossby number $\left(R_{\rm o}^{\rm NL}\right)$. Resonant excitation is obtained for $R_{\rm o}^{\rm NL}\approx R_{\rm o}$. (Taken from \cite{MNT2014}, courtesy {\it Astronomy \& Astrophysics})}
\label{fig1}
\end{center}
\end{figure}

In \cite{MNT2014}, we have formally demonstrated that rotation, through the Coriolis acceleration, modifies the stochastic excitation of gravity waves and GIWs, the control parameters being the wave's Rossby number $R_{\rm o}=\sigma/2\Omega$ ($\sigma$ and $\Omega$ are the wave frequency and the rotation rate respectively) and the non-linear Rossby number $R_{\rm o}^{\rm NL}=t_{\Omega}/t_{\rm c}$ of convective turbulent flows ($t_{\rm c}$ is the convective turn-over time and $t_{\Omega}=\left(2\Omega\right)^{-1}$). On one hand, in the super-inertial regime ($\sigma>2\Omega,\hbox{ {\it i.e.} }R_{\rm o}>1$), the evanescent behaviour of GIWs in convective regions becomes increasingly weaker as the rotation speed grows until $R_{\rm o}=1$. Simultaneously, the turbulent energy cascade towards small scales is slowed down. The coupling between super-inertial GIWs and given turbulent convective flows is then amplified. On the other hand, in the sub-inertial regime ($\sigma<2\Omega,\hbox{ {\it i.e.} }R_{\rm o}<1$), GIWs become propagative inertial waves in convection zones. In the case of rapid rotation, turbulent flows, which become strongly anisotropic, result from their non-linear interactions. Sub-inertial GIWs that correspond to propagating inertial waves in convection zones are then intrinsically and strongly coupled to rapidly rotating convective turbulence. These two different regimes are summarised in Fig. \ref{fig1}.

These results are of great interest for asteroseismic studies since GIW amplitudes are thus expected {to} be larger in rapidly rotating stars, a conclusion that is supported by recent observations and numerical simulations. For example, until recently, stochastic gravity waves were thought to be of too low amplitude to be detected even with space missions such as CoRoT \citep{Samadietal2010}. The discovery of stochastically excited GIWs in the rapid rotator HD\,51452 proved that such waves can be detected. The interpretation of observed pulsational frequencies in rapidly rotating massive stars should take this into account for example in Be and Bn stars. Finally, the transport of angular momentum by stochastic GIWs may have important consequences in active massive stars such as Be stars, in which it may be at the origin of observed outbursts \citep[][]{LNT2014}. 

\bibliographystyle{iau307}
\bibliography{Biblio_Mathis_proc1}

\end{document}